\documentclass[12pt]{article}

\usepackage{graphicx}
\usepackage{amsmath}
\usepackage{amsfonts}
\usepackage{amssymb}

\begin{document}

\renewcommand{\thefootnote}{\fnsymbol{footnote}}

\title{Competitive Dynamics of Web Sites}
\author{Sebastian M. Maurer and Bernardo A. Huberman\footnote{SMM thanks
the Fannie and John Hertz Foundation and the Stanford Graduate
Fellowship for financial support. We thank Eytan Adar and Lada Adamic
for stimulating discussions.}\,\\
Xerox Palo Alto Research Center, Palo Alto, CA 94304}

\maketitle

\begin{abstract}

We present a dynamical model of web site growth in order to explore the
effects of competition among web sites and to determine how they affect
the nature of markets. We show that under general conditions, as the
competition between sites increases, the model exhibits a sudden
transition from a regime in which many sites thrive simultaneously, to a
"winner take all market" in which a few sites grab almost all the users,
while most other sites go nearly extinct. This prediction is in
agreement with recent measurements on the nature of electronic markets.

\end{abstract}

\newpage

\section{Introduction}

The emergence of an information era mediated by the Internet brings
about a number of novel and interesting economic problems. Chiefly among
them is the realization that ever decreasing costs in communication and
computation are making the marginal cost of transmitting and
disseminating information essentially zero. As a result, the standard
formulation of the competitive equilibrium theory is inapplicable to the
Internet economy. This is because the theory of competitive equilibrium
focuses on the dynamics of price adjustments in situations where both
the aggregate supply and demand are a function of the current prices of
the commodities \cite{Hotelling,Arrow}. Since on the Internet the price
of a web page is essentially zero, supply will always match demand, and
the only variable quantity that one needs to consider is the aggregate
demand, i.e. the number of customers willing to visit a site or download
information or software. As we will show, this aggregate demand can
evolve in ways that are quite different from those of price adjustments.

A particular instance of this different formulation of competitive
dynamics is provided by the proliferation of web sites that compete for
the attention and resources of millions of consumers, often at immense
marketing and development costs. As a result, the number of visitors
alone has become a proxy for the success of a web site, the more so in
the case of advertising based business models, where a well defined
price is placed on every single page view. In this case, most customers
do not pay a real price for visiting a web site. The only cost a visitor
incurs is the time spent viewing an ad-banner, but this cost is very low
and practically constant. Equally interesting, visits to a web site are
such that there is non-rival consumption in the sense that one's access
to a site does not depend on other users viewing the same site. This can
be easily understood in terms of Internet economics: once the fixed
development cost of setting up a web site has been paid, it is
relatively inexpensive to increase the capacity the site needs to meet
increased demand. Thus, the supply of served web pages will always track
the demand for web pages (neglecting network congestion issues) and will
be offered at essentially zero cost.

The economics of information goods such as the electronic delivery of
web pages has recently been reviewed by Smith et. al. \cite{Smith}.
They show that when the marginal reproduction cost approaches zero, new
strategies and behaviors appear, in particular with respect to bundling
\cite{Bakos}, price dispersion \cite{Brynjolfsson}, value pricing versus
cost pricing \cite{Varian1}, versioning \cite{Varian2}, and
complicated price schedules \cite{Kephart}.

Since supply matches demand when the price become negligibly small, the
only variable quantity that we will consider in our model is the
aggregate demand, i.e. the number of customers willing to visit a site.
This is the quantity for which we study the dynamics as a function of
the growth and capacity of web sites, as well as the competition between
them. In particular, we explore the effects that competitive pressures
among web sites have on their ability to attract a sizeable fraction of
visitors who can in principle visit a number of equivalent sites. This
is of interest in light of results obtained by Adamic and Huberman
\cite{Adamic}, who showed that the economics of the Internet are such
that the distribution of visitors per site follows a power-law
characteristic of winner-take-all markets. They also proposed a growth
model of the Internet to account for this behavior which invokes either
the continuous appearance of new web sites or different growth rates for
sites.

While such a theory accounts for the dynamics of visits to sites, it
does not take into account actions that sites might take to make
potential visitors to several similar sites favor one over the other. As
we show, when such mechanisms are allowed, the phenomenon of
winner-take-all markets emerges in a rather surprising way, and persists
even in situations where no new sites are continously created.

Our work also explains results obtained from computer simulation of
competition between web
sites by O\u gu\c s et. al. \cite{Ogus}. Their experiments
show that brand loyalty and network effects together result in a form
of winner-take-all market, in which only a few sites survive. This is
consistent with the predictions of our theory.

In section 2 we present the model and illustrate its main predictions by
solving the equations in their simplest instance in section 3. In
section 4 we show that the transition from fair market share to
winner-take-all persists in the general case of competition between two
sites and in section 5 we extend our results to very many sites. We also
show the appearence of complicated cycles and chaotic outcomes when the
values of the competitive parameters are close to the transition point. A
concluding section summarizes our results and discusses their implications
to electronic commerce.

\section{The Model}

Consider $n$ web sites offering similar services and competing for the
same population of users, which we'll take to be much larger than the
number of sites. Each site engages in policies, from advertising to
prize reductions, that try to increase their share of the customer base
$f_i$. Note that while $f_i$ is the fraction of the population that is a
customer of web site $i$, it can be more generally taken to be the
fraction of the population aware of the site's existence. This could be
measured by considering the number of people who bookmark a particular
site.

The time evolution of the customer fraction $f_i$ at a given site $i$ is
determined by two main factors. If there is no competition with any
other sites, the user base initially grows exponentially fast, at a
rate $\alpha_i$, and then saturates at a value $\beta_i$. These values are
determined by the site's capacity to handle a given number of visitors per
unit time. If, on the
other hand, other sites offer competing services, the strength of the
competition determines whether the user will be likely to visit
several competing sites (low competition levels) or whether having
visited a given site reduces the probability of visiting another (high
competition level).

Specifically, the competition term can be understood as follows: if
fractions $f_i$ and $f_j$ of the people use sites $i$ and $j$,
respectively, then assuming that the probability of using one site is
independent of using another, a fraction $f_i f_j$ will be using both
sites. However, if both sites provide similar services, then some of
these users will stop using one or the other site. The rate at which
they will stop using site $i$ is given by $\gamma_{ij} f_i f_j$, and the
rate at which they abandon site $j$ is given by $\gamma_{ji} f_i f_j$
(note that $\gamma_{ij}$ is not necessarily equal to $\gamma_{ji}$).

Mathematically the dynamics can thus be expressed as

\begin{eqnarray}
\frac{{\rm d}f_i}{{\rm d}t} & = &
\alpha_{i} f_i (\beta_i - f_i) - \sum_{i \ne j} \gamma_{ij} f_i f_j,
\label{eq1}
\end{eqnarray}

\noindent where $\alpha_i$ is the growth rate of individual sites,
$\beta_i$ denotes their capacity to service a fraction of the customer
base and $\gamma_{ij}$ is the strength of the competition. The
parameter values are such that $\alpha_i \geq 0$, $0 \leq \beta_i \leq 1$
and
$\gamma_{ij} \geq 0$.

The system of equations (\ref{eq1}), which determines the nonlinear
dynamics of user visits to web sites, possesses a number of attractors
whose stability properties we will explore in detail \footnote{The
equations are functionally similar to those describing the competition
between modes in a laser \cite{Goel}, and to those describing
prey-predator equations in ecology \cite{Murray}.}. In particular, we
will show that as a function of the competition level, the solutions can
undergo bifurcations which render a particular equilibrium unstable and
lead to the appearance of new equilibria. The most striking result among
them is the sudden appearance of a winner-take-all site which captures
most of the visitors, a phenomenon that has been empirically observed in
a study of markets in the web \cite{Adamic}.

Since the complexity of the equations may obscure some the salient
features of the solutions, we will first concentrate on the simplest
case exhibiting a sharp transition from fair market share to a
winner-take-all site, and then consider more complicated examples.

\section{Fair Market Share to Winner-Take-All}

Let us first consider one of the simplest instances of the problem described
above, in which two web sites have the same growth rates
$\alpha_1 = \alpha_2 = 1$, the same capacities $\beta_1 = \beta_2 = 1$
and symmetric competion $\gamma_{12} = \gamma_{21} = \gamma$. In this
case the equations take the form

\begin{eqnarray}
\frac{{\rm d}f_1}{{\rm d}t} & = & f_1 (1 - f_1 - \gamma f_2) \nonumber \\
\frac{{\rm d}f_2}{{\rm d}t} & = & f_2 (1 - f_2 - \gamma f_1) \nonumber
\end{eqnarray}

The four fixed points of this equation, which determine the possible
equilibria, are given by \begin{displaymath} (f_1, f_2) \in
\{(0, 0), (1,0), (0, 1), (\frac{1}{1 + \gamma},\frac{1}{1 + \gamma})\}
\nonumber \end{displaymath}

Since not all of these equilibria are stable under small perturbations,
we need to determine their time evolution when subjected to a sudden
small change in the fraction of visitors to any site. To do this, we need to
compute the eigenvalues of the Jacobian
evaluated at each of the four fixed points. The Jacobian is

\begin{displaymath}
\mathbf{J} = \left( \begin{array}{cc} 1 - 2 f_1^0 - \gamma f_2^0 &
- \gamma f_1^0 \\ - \gamma f_2^0 & 1 - 2 f_2^0 - \gamma f_1^0
\end{array} \right) \nonumber
\end{displaymath}

\noindent and the eigenvalues at each of the fixed points are given in
the following table:

\begin{center}
\begin{tabular}{cc} equilibrium & eigenvalues \\
\hline \hline
$(0, 0)$ & 1 (twice) \\
$(\frac{1}{1 + \gamma},\frac{1}{1 + \gamma})$ &
$-1$ and $\frac{\gamma - 1}{1 + \gamma}$ \\
$(1, 0)$ or $(0,1)$ & $\frac{1}{2}(- \gamma \pm \sqrt{(2 - \gamma)^2})$
\end{tabular}
\end{center}

\begin{figure}[tbh] \begin{center} \includegraphics[scale=0.5]{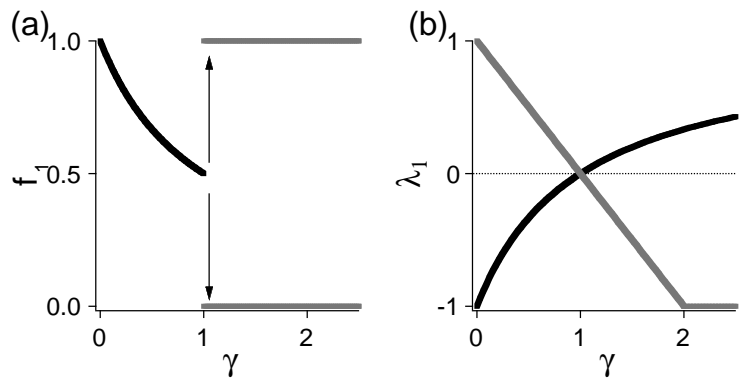}
\end{center} \caption{(a) Equilibrium values for $f_1$ and (b) largest
eigenvalues of the Jacobian for the $(\frac{1}{1 + \gamma},\frac{1}{1 +
\gamma})$ fixed point (dark) and the $(1,0)$ fixed points (light), as a
function of the competition value $\gamma$.}
\label{fig1}
\end{figure}

  From this it follows that the fixed point $(0, 0)$ is never stable. On
the other hand, the equilibrium at $(\frac{1}{1 + \gamma},\frac{1}{1 +
\gamma})$ is stable provided that $\gamma < 1$. And the fixed points
$(0,1)$ and $(1,0)$ are both stable if $\gamma > 1$. From these results,
we can plot the equilibrium size of the customer population as a
function of the competition $\gamma$ between the two competitors. As
Figure \ref{fig1} shows, there is a sudden, discontinuous transition at
$\gamma = 1$. For low competition, the only stable configuration has
both competitors sharing the market equally. For high competition, the
market transitions into a "Winner-Take-All Market" \cite{Frank}, in
which one competitor grabs all the market share, whereas the other
gets nothing.

As we will show below, this sudden transition persists under extremely
general conditions, for two competitors as well as for $n$ competitors.
The significant feature is that a very small change in the parameters
can radically affect the qualitative nature of the equilibrium.

Another feature is that near the transition, the largest eigenvalue of
the stable state is very close to zero (but negative). This means that
the time of convergence to equilibrium diverges. In more complicated
systems, this may make it extremely difficult to predict which
equilibrium the system will converge to in the long term.

\section{Competition between two sites}

We now analyse the two site model in its most general form, without
restricting the values of the parameters to be the same for the two
sites. The system of equations is

\begin{eqnarray}
\frac{{\rm d}f_1}{{\rm d}t} & = &
f_1 (\alpha_1 (\beta_1 - f_1) - \gamma_{12} f_2) \nonumber \\
\frac{{\rm d}f_2}{{\rm d}t} & = &
f_2 (\alpha_2 (\beta_2 - f_2) - \gamma_{21} f_1) \nonumber
\end{eqnarray}

As in the previous section these equations possess four fixed points at:

\begin{eqnarray}
(f_1^0, f_2^0) & = & (0, 0) \nonumber \\
(f_1^0, f_2^0) & = & (\beta_1, 0) \nonumber \\
(f_1^0, f_2^0) & = & (0, \beta_2) \nonumber \\
(f_1^0, f_2^0) & = &
(\frac{\alpha_2 (\alpha_1 \beta_1 - \beta_2 \gamma_{12})}
          {\alpha_1 \alpha_2 - \gamma_{12} \gamma_{21}},
     \frac{\alpha_1 (\alpha_2 \beta_2 - \beta_1 \gamma_{21})}
          {\alpha_1 \alpha_2 - \gamma_{12} \gamma_{21}}) \nonumber
\end{eqnarray}

Let's analyze the stability of each fixed point. This is done by
evaluating the Jacobian

\begin{displaymath}
\mathbf{J} = \left( \begin{array}{cc}
\alpha_1 \beta_1 - 2 \alpha_1 f_1^0 - \gamma_{12} f_2^0 &
- \gamma_{12} f_1^0 \\
- \gamma_{21} f_2^0 &
\alpha_2 \beta_2 - 2 \alpha_2 f_2^0 - \gamma_{21} f_1^0
\end{array} \right) \nonumber
\end{displaymath}

\noindent and computing its two eigenvalues. Each fixed point will be
stable only
if the real parts of both eigenvalues are negative. The first trivial
fixed point is always unstable, since the eigenvalues $\alpha_1
\beta_1$ and $\alpha_2 \beta_2$ are both positive
quantities. The second equilibrium, with $(f_1, f_2) = (\beta_1, 0)$ is
stable provided $\frac{\gamma_{12}}{\alpha_1} >
\frac{\beta_1}{\beta_2}$. The third equilibrium $(f_1, f_2) = (0,
\beta_2)$ is similarly stable only if $ \frac{\gamma_{21}}{\alpha_2} >
\frac{\beta_2}{\beta_1}$. The final case is the most complicated one,
and is stable in three distinct regimes. However, two of the stable
solutions have a negative $f_1$ or $f_2$ and can never be reached from
an initial condition with both populations positive. The only remaining
equilibrium is stable whenever the other two aren't. In order to
summarize these results in the table below, it is convenient to define
the following parameters:

\begin{displaymath}
\alpha = \frac{\alpha_1}{\alpha_2} \qquad
\beta = \frac{\beta_1}{\beta_2} \qquad
\gamma_1 = \frac{\gamma_{12}}{\alpha_1} \qquad
\gamma_2 = \frac{\gamma_{21}}{\alpha_2}. \nonumber
\end{displaymath}

\begin{center}
\begin{tabular}{cc} equilibrium & stable if \\
\hline
\hline
(0, 0) & never \\
$(\beta_1, 0)$ & $\beta > \frac{1}{\gamma_2}$ \\
$(0, \beta_2)$ & $\beta < \gamma_1$ \\
$(\frac{\beta_1 - \beta_2 \gamma_{1}} {1 - \gamma_{1} \gamma_{2}},
      \frac{\beta_2 - \beta_1 \gamma_{2}}{1 - \gamma_{1} \gamma_{2}})$ &
$\begin{array}{cc}
\beta > \gamma_1 & \beta < \frac{1}{\gamma_2}
\end{array}$ \\
\end{tabular}
\end{center}

\begin{figure}[tbh] \begin{center} \includegraphics[scale=0.5]{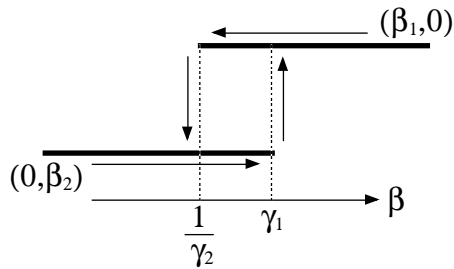}
\end{center} \caption{Hysteresis behavior as $\beta$ is changed with
$\gamma_1$ and $\gamma_2$ fixed.}
\label{fig2}
\end{figure}

Note that in the last row, $\beta > \gamma_1$ and $\beta <
\frac{1}{\gamma_2}$ imply that $\gamma_1 \gamma_2 < 1$. As a result, for
fixed $\gamma_1$ and $\gamma_2$, there are two different regimes. Either
$\gamma_1 \gamma_2 < 1$, in which case intermediate values of $\beta$
($\gamma_1 < \beta < \frac{1}{\gamma_2}$) lead to a "fair" equilibrium
in which both sites get a non zero $f_i$, or $\gamma_1 \gamma_2 < 1$, in
which case intermediate values of $\beta$ ($\frac{1}{\gamma_2} < \beta <
\gamma_1$) lead to a situation in which either of the two
"winner-take-all" equilibria is stable. In this latter case, hysteresis
occurs if $\beta$ slowly changes with time. This is illustrated in
Figure \ref{fig2}. Starting from a low value of $\beta$, the only stable
equilibrium is $(0, \beta_2)$. If we slowly increase the ratio $\beta$,
this equilibrium remains stable, until $\beta > \gamma_1$. At that point,
the equilibrium $(0, \beta_2)$ becomes unstable and the system relaxes to
the only new equilibrium $(\beta_1, 0)$. If we now reverse the process,
and decrease the value of $\beta$, the new solution remains stable as long
as $\beta > \frac{1}{\gamma_1}$.

Thus, there always is at least one stable equilibrium, but there never
are more than two. If there are two stable equilibria, then the initial
conditions determine into which of the two the system will fall. Which
of the equilibria are stable depends on a total of three parameters
(down from the five parameters that are required to fully describe the
system once the time variable is rescaled).

\begin{figure}[tbh] \begin{center}
\includegraphics[scale=0.5]{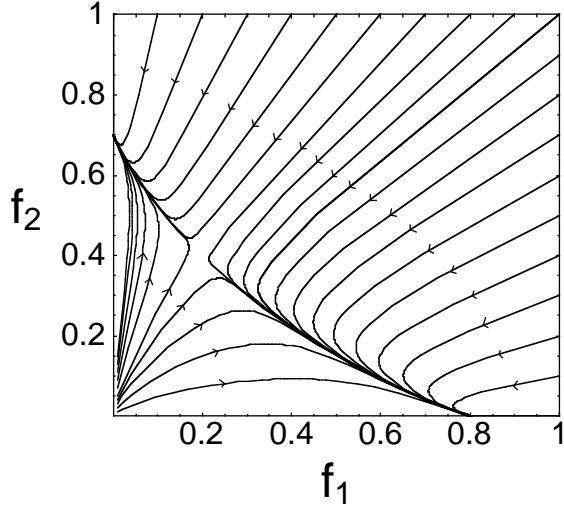}
\end{center} \caption{Basins of attraction for the solutions to a
winner-take-all market as a function of the initial market share, for
$\alpha_1 = 1$, $\alpha_2 = 0.8$, $\beta_1 = 0.8$, $\beta_2 = 0.7$,
$\gamma_{12} = 1.5$ and $\gamma_{21} = 1.2$}
\label{fig3}
\end{figure}

It is interesting to ask which of the two sites will win as a function
of a set of fixed parameters and as a function of the starting point.
To do this, we plot the motion of $(f_1, f_2)$ as a vector field in
Figure \ref{fig3}, for a particular set of parameters. As can be seen,
the space of initial conditions is divided into two distinct regions,
each of which leads to a different equilibrium.

\section{Many sites}

\subsection{Analytic treatment}

We now show that the sharp transition to a winner-take-all market that
we found in the two site case is also present when many sites are in
competition. In order to do so, we first examine the case where the
parameters are the same for all sites, $i$, so that Equation (\ref{eq1}) can
be rewritten as

\begin{eqnarray}
\frac{{\rm d}f_i}{{\rm d}t} & = &
f_i (\alpha \beta - \alpha f_i - \gamma \sum_{i \ne j} f_j). \nonumber
\end{eqnarray}

\noindent where $i = 1, \ldots, n$ and $n$ is the number of sites. For
$n$ equations, there are $2^n$ different vectors $(f_1,\ldots,f_n)$ for
which all the time derivatives are zero, since for each equation, either
$f_i = 0$ or $\alpha \beta - \alpha f_i - \sum_{i \ne j} \gamma f_j = 0$
at equilibrium. Without loss of generality, we can relabel the $f_i$
such that the first $k$ of them are non-zero, while the remaining $n$
are zero.

At equilibrium, the value of the $f_i$ with $1 \leq i \leq k$ will be
given by the solution of

\begin{eqnarray}
\left( \begin{array}{cccc}
\alpha & \gamma & \ldots & \gamma \\
\gamma & \alpha & \ldots & \gamma \\
\vdots & \vdots & \ddots & \vdots \\
\gamma & \gamma & \ldots & \alpha \end{array} \right)
\left( \begin{array}{c}
f_1 \\
f_2 \\
\vdots \\
f_k \end{array} \right)
=
\left( \begin{array}{c}
\alpha \beta \\
\alpha \beta \\
\vdots \\
\alpha \beta
\end{array} \right) \nonumber
\end{eqnarray}

Except for the degenerate cases, the matrix on the left hand side is
invertible, so that

\begin{eqnarray}
f_i = \left\{ \begin{array}{cl}
\frac{\alpha \beta}{\alpha + (k - 1) \gamma} &
\textrm{if $1 \leq i \leq k$} \\
0 & \textrm{if $k + 1 \leq i \leq n$} \end{array} \right. \nonumber
\end{eqnarray}

We are now ready to compute the Jacobian about this equilibrium. It
takes the form

\begin{eqnarray}
\mathbf{J} = \left( \begin{array}{ccccccc}
X & Y & \ldots & Y & 0 & \ldots & 0 \\
Y & X & \ldots & Y & 0 & \ldots & 0 \\
\vdots & \vdots & \ddots & \vdots & \vdots & & \vdots \\
Y & Y & \ldots & X & 0 & \ldots & 0 \\
0 & 0 & \ldots & 0 & Z & \ldots & 0 \\
\vdots & \vdots & & \vdots & \vdots & \ddots & \vdots \\
0 & 0 & \ldots & 0 & 0 & \ldots & Z \end{array} \right) \nonumber
\end{eqnarray}

\noindent where

\begin{eqnarray}
X & = & \alpha \beta - (2 \alpha + (k - 1) \gamma)
\frac{\alpha \beta}{\alpha + (k - 1) \gamma} \nonumber \\
Y & = &
- \gamma \frac{\alpha \beta}{\alpha + (k - 1) \gamma} \nonumber \\
Z & = &
\alpha \beta - k \gamma \frac{\alpha \beta}{\alpha + (k - 1) \gamma}
\nonumber
\end{eqnarray}

The eigenvalues of the Jacobian are $Z$ (with multiplicity $n - k$), $X
- Y$ (with multiplicity $k - 1$) and $X + (k - 1) Y$ (with multiplicity
$1$). Note that the last two eigenvalues are absent if $k = 0$. Thus
there are four distinct cases to check: $k = 0$, $k = 1$, $1 < k < n$
and $k = n$. In the first case, the only eigenvalues are $Z = \alpha
\beta > 0$, so this solution is always unstable.

With $k = 1$, the eigenvalues are

\begin{eqnarray}
X & = & - \alpha \beta \nonumber \\
Z & = & (\alpha - \gamma) \beta \nonumber
\end{eqnarray}

That is, an equilibrium with one out of $n$ winners is stable provided
$\alpha < \gamma$. Note that this is the same condition we obtained for
two competitors.

With $1 < k < n$ we have the eigenvalues

\begin{eqnarray}
Z & = &
\alpha \beta \frac{\alpha - \gamma}{\alpha + (k - 1) \gamma} \nonumber \\
X - Y & = &
\alpha \beta \frac{\gamma - \alpha}{\alpha + (k - 1) \gamma} \nonumber \\
X + (k - 1) Y & = & - \alpha \beta \nonumber
\end{eqnarray}

The first and second eigenvalues above can not be negative
simultaneously, thus there are no stable solutions with $1 < k < n$.

Finally, for $k = n$ we have the same eigenvalues as for $1 < k < n$,
except that the eigenvalue $Z$ is now inexistent. As a result, the
solution with $k = n$ is stable provided that $\gamma < \alpha$.

To summarize, the only stable solutions are

\begin{eqnarray}
f_i = \left\{ \begin{array}{cl}
\frac{\alpha \beta}{\alpha + (n - 1) \gamma} &
\textrm{if $\gamma < 1$} \\
\beta \delta_{ij} & \textrm{if $\gamma > 1$} \end{array} \right.
\nonumber
\end{eqnarray}

That is, the winner-take-all dynamics observed for two sites persists
independently of the number of competitors involved, at least in an
idealized symmetric configuration. In the next section, we consider the
dynamics for large systems in which the parameter value are drawn from a
random
distribution.

\subsection{Critical dynamics}

In the most general case, the dynamical change in the fraction of
visitors to web sites can be determined by numerically solving the
general equations of our model. In addition, provided the number of
sites $n$ remains small one can check each of the $2^n$ candidate
equilibria for stability and verify whether the numerical simulation
converged to the only equilibrium or missed an existing but hard to
reach equilibrium.

\begin{figure}[tbh] \begin{center} \includegraphics[scale=0.5]{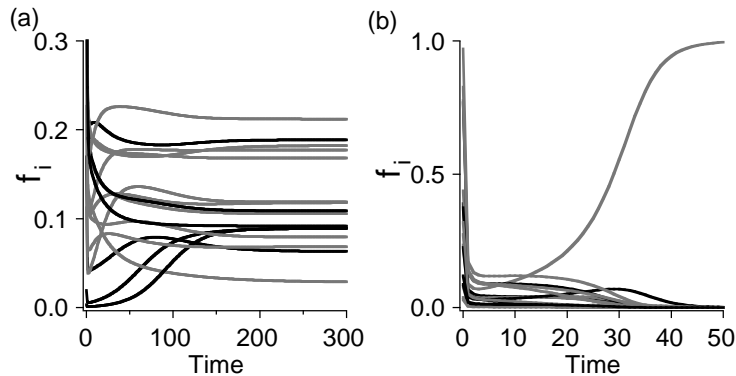}
\end{center} \caption{Solutions obtained by numerical integration for
two different distributions of parameters. In (a) we are below the
transition to winner-take-all, with $\bar{\gamma} = 0.5$,
while in (b) we are above the transition to winner-take-all, with
$\bar{\gamma} = 1.5$. In both cases $\alpha_i = \beta_i = 1$ for
all $i$.} \label{fig4} \end{figure}

In Figure \ref{fig4}, we show the time evolution of $f_i$ for sixteen
web sites, obtained by numerically integrating the equations using a
Runge-Kutta scheme. The parameters defining the competitive strength
between sites, $\gamma_{ij}$, were randomly chosen from a Gaussian
distribution with a standard deviation of $0.1$, and a fixed mean
$\bar{\gamma}$. On the left panel we exhibit a solution for
$\bar{\gamma} = 0.5$, far below the transition point. On the right
panel, $\bar{\gamma} = 1.5$ places us well above the
transition, and we observe the evolution towards a winner-take-all market.
Whereas below the transition the equilibrium has all sixteen competitors
sharing the market, above it one web site takes all visitors.

Given the fixed set of parameters for the model, it is possible to
diagonalize the Jacobian, evaluated at each of the $2^n = 2^{16}$ fixed
points. As in the case of the symmetric case, or for the general two
site case, only one equilibrium is stable when $\bar{\gamma} \ll 1$. In
addition, the values of the $f_i$ at the single stable equilibrium found
in this manner match the values that the numerical simulation converges
to.

When $\bar{\gamma} \gg 1$, numerically diagonalizing the $2^n$ Jacobians
shows that the $n$ equilibria of the form $f_i = \delta_{ik}$, and no
others, are stable. Thus the transition to a winner-take-all market
subsists even when the parameters come from a randomized distribution.

A more interesting situation is posed by the dynamics of competition
when the competitive strength approaches the critical value,
$\bar{\gamma} \approx 1$. Since near the transition point the largest
eigenvalue has an absolute value very close to zero the transients to
equilibrium are very long. Moreover, the nature of the transients is
such that many sites alternate in their market dominance for long
periods of time.

\begin{figure}[tbh] \begin{center}
\includegraphics[scale=0.5]{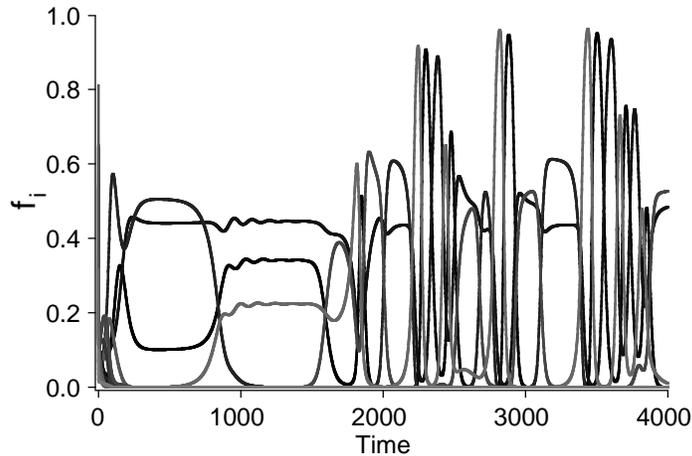}
\end{center}
\caption{Dynamics near the transition to winner-take-all, for the same
parameters as in Figure \ref{fig4}, but with $\bar{\gamma} = 1.0$.}
\label{fig5}
\end{figure}

Numerical diagonalization of all the possible Jacobians near criticality
shows that frequently there are several stable equilibria, in which some
sites have non-zero $f_i$, and some sites don't. However, these
solutions typically are not reached in a finite amount of time (if at all)
when numerically integrating the equations. Furthermore, numerical
integration, as shown in Figure \ref{fig5}, suggests that for this range
of parameters the dynamics are chaotic (small differences in the initial
conditions lead to diverging trajectories). For some initial conditions
the system may converge to limit cycles, rather that to static
equilibria. Thus, when the parameter values of $\gamma_{ij}$ are
drawn from a distribution, the transition is not sudden in $\bar{\gamma}$.
There is a range of values of $\bar{\gamma}$ for which the dynamics are
more complicated.

For much larger $n$, it is no longer possible to verify every single
candidate fixed point for stability. However, it is still possible to
numerically integrate the equations. Either way, as the example in
Figure \ref{fig5} shows, the question of existence and stability of the
equilibria is irrelevant if the stable equilibria are never reached, or
only reached after an unreasonable amount of time.

\section{Conclusion}

In this paper we have shown that under general conditions, as the
competition between web sites increases, there is a sudden transition
from a regime in which many sites thrive simultaneously, to a "winner
take all market" in which a few sites grab almost all the users, while
most other sites go nearly extinct, in agreement with the observed
nature of electronic markets. This transition is the result of a
nonlinear interaction among sites which effectively reduces the growth
rate of a given site due to competitive pressures from the others.
Without the interaction term, web sites would grow exponentially fast to a
saturation level that depends on their characteristic properties.

Moreover, we have shown that the transition into a winner-take-all
market occurs under very general conditions and for very many sites. In
the limiting case of two sites, the phenomenon is reminiscent of the
"Principle of Mutual Exclusion" in ecology \cite{Murray,Hsu,Pianka}, in
which two predators of the same prey can not coexist in equilibrium when
competitive predation is very strong.

Smith et. al. \cite{Smith,Brynjolfsson} attribute the price dispertion
of goods sold online to several features of web sites: differences in
branding and trust, in the appearance and in the quality of the search
tools, switching costs between sites, and last but not least retailer
awareness. A winner-take-all economy may thus have strong consequences
for price dispertion, since a few sites can charge more by virtue of
dominating the mind share of their customers.

It is interesting to speculate about the applicability of this model to
different markets. We motivated the model for a massless Internet
economy, in which demand can be instantly satisfied by supply at a
negligible cost to the supplier, and in which competition does not occur
on the basis of cost, but rather on advertising and differentiation in
the services provided by the web sites. However, since winner-take-all
markets
are being observed in a much broader range of markets, it might well be
the case that the sudden
transition to winner-take-all behavior might also be a feature of these
markets as well.

\end{document}